# Using network science to analyze football passing networks: dynamics, space, time and the multilayer nature of the game.

J.M. Buldu[1,2,3], J. Busquets[4], J.H. Martínez[3,5], J.L. Herrera-Diestra[6], I. Echegoyen[1,2,3], J. Galeano[7] and J. Luque[8]

[1] Laboratory of Biological Networks, Center for Biomedical Technology, UPM, Madrid, Spain
[2] Complex Systems Group & GISC, Universidad Rey Juan Carlos, Móstoles, Spain
[3] Grupo Interdisciplinar de Sistemas Complejos (GISC), Madrid, Spain
[4] ESADE Business School, Barcelona, Spain
[5] INSERM-Institute du Cerveau et de la Moelle Épinière. H. Salpêtrière, Paris, France
[6] ICTP South American Institute for Fundamental Research, IFT-UNESP, São Paulo, Brazil
[7] Grupo de Sistemas Complejos, Universidad Politécnica de Madrid, Madrid, Spain
[8] Telefónica Research, Barcelona, Spain

## *Introduction*

During the last decade, *Network Science* has become one of the most active fields in applied physics and mathematics, due to the fact that there is a diversity of social, biological and technological systems that can be analyzed under the scope of this, relatively new, branch of science (Newman, 2010). Network Science analyzes systems consisting of a number of coupled "units" and focuses on the influence that network structure has on the processes occurring in the whole system, combining tools coming from four different fields: graph theory, statistical physics, nonlinear dynamics and Big Data (Barabási, 2016). The power of Network Science relies on the fact that, once a system is composed of a series of interacting units, it can be projected into a network, i.e., a set of $N$ nodes connected through $L$ links, for further analyses, no matter what the nature of the units and their interactions is. In this way, *Network Science* has given a new perspective about problems as different as, for example, (i) the structure of interactions of social networks (e.g., Facebook, Twitter,…) and how it hinders/enhances the transmission of information/rumours (Ferrara, 2012; Grabowicz et al., 2012), (ii) how brain regions are coordinated to carry out a cognitive or motor task (Sporns and Bullmore, 2009; Papo et al., 2014) or (iii) what is the robustness of a power grid when a certain number of cascading failures occur (Brummitt et al., 2012).

From the diversity of applications of Network Science, in this Opinion paper we are concerned about its potential to analyze one of the most extended group sports: Football (soccer in U.S. terminology) (Sumpter, 2016). As we will see, Network Science allows addressing different aspects of the team organization and performance not captured by classical analyses based on the performance of individual players. The reason behind relies on the complex nature of the game, which, paraphrasing the foundational paradigm of complexity sciences "*can not be analyzed by looking at its components (i.e., players) individually but, on the contrary, considering the system as a whole*" or, in the classical words of after-match interviews *"it's not just me, it's the team"*.

The recent ability of obtaining datasets of all events occurring during a match, including the position of the players and the interactions between them has opened the door to new kind of studies where it is possible to analyse and quantify the behaviour of a team as a whole, together with the role of a single player (Gudmundsson and Horton, 2017). Under this framework, the organization of a team can be considered as the result of the interaction between its players, creating a network based on passes. In this way, we can create passing *networks,* which are directed (i.e., links between players go in one direction), weighted (i.e., the strength of the links is based on the number of passes between players), spatially embedded (i.e., the Euclidean position of the ball and players is highly relevant) and time evolving (i.e., the network continuously changes its structure).

## *Passing networks: Information from a new perspective*

The seminal paper by Gould and Gatrell (1977) introducing the concept of passing networks in a football match didn't obtain the relevance it deserved, both in the scientific and sports communities. However, some decades later, the work of Duch et al. (Duch et al., 2010) marked the start of a decade, the current one, that is witnessing how the analysis of passing networks unveils crucial information about the organization, evolution and performance of football teams and players. Passes along the match give rise to three main types of passing networks: (i) *player passing networks*, where nodes are the players of a team (Passos et al., 2011; Grund, 2012), (ii) *pitch passing networks*, where nodes are specific regions of the field connected through passes made by players occupying them (Cintia et al., 2015) or (iii) *pitch-player passing networks*, where nodes are a combination of a player and its position at the moment of the pass (Narizuka et al., 2014; Cotta et al., 2013). In all cases, the precise time and position of all passes along the match is translated into a network for a sub-sequent analyses.

Once the network is constructed, several "*topological scales*" (i.e., scales at which the network organization is analyzed) can be identified inside the passing network of a football team: (i) the *microscale*, where the analysis is carried out at the level of nodes, i.e. the players and its role inside the network, (ii) the *mesoscale*, which ranges from small motifs describing the interaction of 2 or 3 players to the detection of larger groups of players that interact most frequently between them and (iii) the *macroscale*, which considers the network as a whole. At each scale, it is possible to obtain a series of *metrics* related to the organization of the team and analyze the interplay between the network metrics and the team (or player) performance.

At the topological microscale, the importance of each player has been related to: its *degree*, which is the number of passes made by a player (Cotta et al., 2013); *eigenvector centrality*, a measure of importance obtained from the eigenvectors of the adjacency matrix (Cotta et al., 2013); *closeness*, measuring the minimum number of steps that the ball has to undergo from one player to reach any other in the team (López-Peña et al., 2012); or *betweenness centrality*, which accounts how many times a given player is necessary for completing the routes (made by the ball) connecting any other two players of its team (Duch et al., 2010; López-Peña et al., 2012). Other metrics, such as the *clustering coefficient*, which measures the number of "neighbours" of a player that also have passed the ball between them (i.e., the number of triangles around a player), has also been quantified to evaluate the contribution of a given player to the local robustness of the passing network (López-Peña et al., 2012).

At the mesoscale level, the analysis of network motifs has shown how the overabundance of certain kinds of passes between groups of three/four players can be related to both the success of team (Gyarmati et al., 2014) and the identification of leaders in the passing network (López-Peña and Sánchez-Navarro, 2015). Concerning the existence of communities of players playing tightly connected between them, Clemente et al. (2015), related the high heterogeneity of the number of passes between players to the existence of sub-communities, which would hinder the behaviour of the team as a whole.

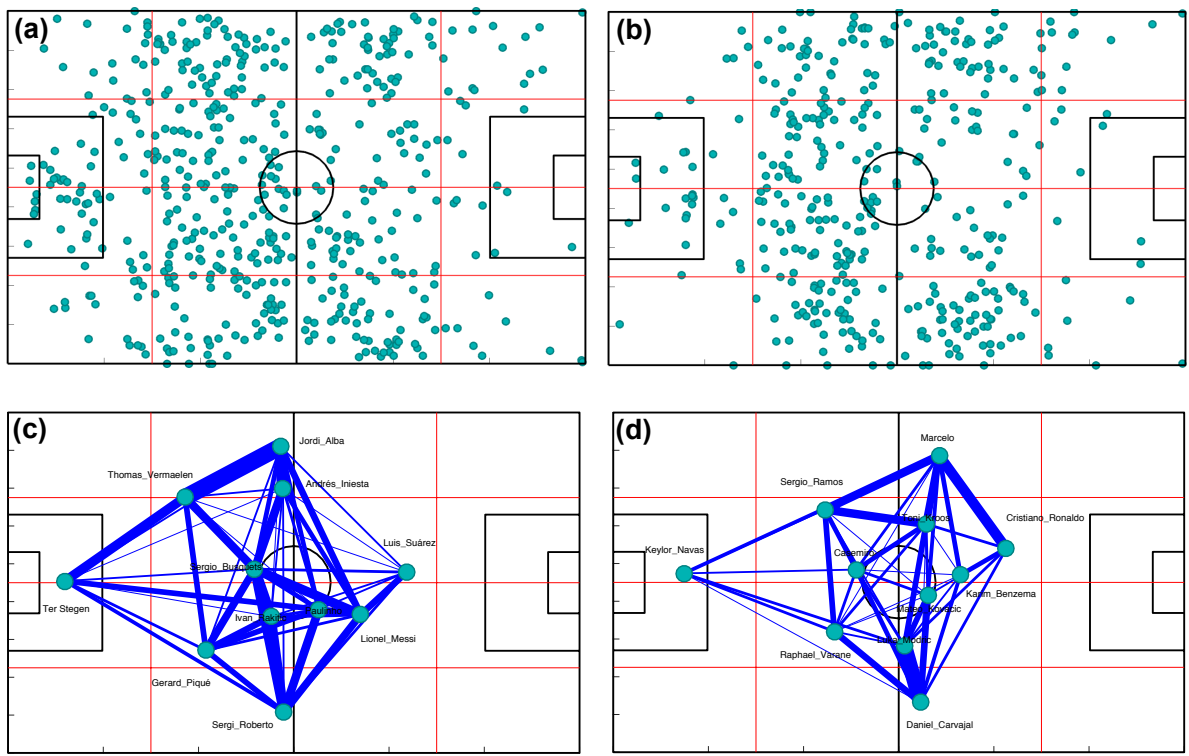

**Fig. 1.-** Construction of a passing network. In this example, passes from the match Real Madrid –Barcelona of the Spanish national league "*La Liga*", season 2017/2018. In the upper row, initial position of all passes made by Barcelona (a) and Real Madrid (b). In the bottom row, Barcelona (c) and Real Madrid (d) passing networks, where link widths are proportional to the number of passes between players, whose position in the network is given by the average position of all their passes. Data provided by Opta.

Finally, at the topological macroscale, a diversity of network metrics has been shown to be informative about the style and performance of football teams. For example, the position of the network centroid, which consists on the center of mass of the passing networks, has been related to the performance of the teams (the more forward, the better) and has been shown to move backwards when teams play as visitors (Bialkowski et al., 2014). Other positional variables, such as the stretch index (mean dispersion of the players around the centroid), the surface area or the team length and width have also been used as more sophisticated metrics related to team performance (Duarte et al., 2012). On the other hand, Duch et al., (2010) designed a performance metric based on the betweenness of the players, showing how it correlated with the probability of winning a match. Other macro-scale measures such as the team average degree (i.e., average number of passes) or the variability of the players' degrees have also been identified as proxies for evaluating team performance (Cintia et al., 2015; Pina et al., 2017). Concerning the macro-scale topology of the passing network, the small-world property (Watts and Strogatz, 1998), observed in a diversity of social, biological and technological networks, has also been reported, suggesting that the reduction of the number of steps to go from a player to any other in the team is promoted along the match (Narizuka et al., 2014). The average clustering coefficient of the team has also been shown to be much higher than equivalent random networks, unveiling the creation of triplets between players (Cotta et al., 2013). More recent studies tracking the position of the players have shown that it is better to maintain a balanced betweenness and a high closeness along the nodes of the passing network (Gonçalves et al., 2017).

### *Challenges: dynamics, space, time and interaction between teams*

Passing networks are, in fact, dynamical system themselves, whose state along time can only be described when considering all the variables that influence their evolution. However, the full identification and quantification of how variables determine the evolution of the game of a team is still an open problem. Indeed, the problem goes beyond the identification of all variables, since the game cannot escape from the existence of stochastic forces that, combined with the high complexity of its intrinsic dynamics, make the modelling and forecasting of a football match a highly challenging task. Distinguishing noise from determinism is an issue where Network Science can help, since it is possible to determine the *level of randomness* of the topology of the network and the dynamics occurring in it (e.g., how the ball moves along the network). In this way, we are still on the way of constructing adequate null models of passing networks that are able to quantify the amount of disorder and complexity of the network topology. These null models must be as realistic as possible and include the intrinsic features of the game such as the degree distribution, length of the passes and realistic positions of the players in the field. Concerning the dynamics along the network, recent approaches using Markovian models could be a starting point to unravel hidden patterns in the passing sequences of a team (López-Peña, 2014) but must include the particular features of player's movements and ability of decision, which are far from being random.

Nonetheless, topology is only one dimension of the analysis of passing networks and at least other two must be included to have a complete picture of the game: space and time. Concerning the former, the division of the pitch into different sub-regions has been carried out at a series of papers, however, it is not clear what is the most adequate partition of the field or, even if a unique partition (or scale) exists for a given team. From the division of the pitch into *6x3* rectangles of equal size up to a segmentation of 100 regions (Cintia et al., 2015) a diversity of field partitions has been suggested (Camerino et al., 2012; Narizuka et al., 2014; Arriaza-Ardiles et al., 2018). The translation of passes into pitch or pitch-player networks seems to be a promising complementary vision of the game, despite the information loss about the players' behaviour.

Time is another dimension traditionally overlooked when constructing passing networks. Note that the analysis of passing networks must take into account its continuous evolution in time and space. It is not enough to describe its state at a certain moment of the match, or even its average. For example, as shown in Duarte et al. (2012), entropy decreases as the game evolves, which is attributed to the fatigue of the players. In addition, collective behaviours decrease in complexity/irregularity during the time periods of the match, accompanied with an increase in the magnitude of deviations from the mean tendency. Network's density, heterogeneity and centralization show differences between the 1st and 2nd part of the matches (Clemente et al., 2015). However, it is common to average along the whole match (Gonçalves et al., 2017), or even along a competition (López-Peña and Touchette, 2012; Gyarmati et al., 2014), obtaining a network that can be informative about the general behaviour of a team but excludes the unavoidable fluctuations that occur during a match. While considering the networks during each of the two parts (Clemente et al., 2015) or constructing sliding windows of a certain length (between 5 and 15 minutes) are reasonable approaches (Yamamoto et al., 2011; Duarte et al., 2012; Cotta et al., 2013), the fact that a team may change its organization along certain periods of different length (Wei et al., 2013) and, moreover, how to detect a change of this organization, increases the complexity of the analysis, suggesting the adaptation of classical mesoscale detection algorithms (Fortunato, 2010) applied to the temporal fluctuations the passing network.

On the other hand, a football match is the result of the competition between two teams, i.e. the interaction between two networks. Therefore, the passing network of a team must be analyzed in combination with the network of the opponent (Narizuka et al., 2014). In this way, it will be possible to draw conclusions about how a team adapts its game depending on the opponent and what kind of topological organization leads to better results. Recent studies about *network-of-networks* in other fields have shown that when networks get connected, important properties of the ensembled systems are modified (Kivelä et al, 2014; Boccaletti et al., 2014). With this regard, a multilayer description of a match, with two interacting layers composed of the internal passes of each team, is still missing. However, this could be a fundamental approach to understand the evolution and adaptability of the teams along the match, which cannot be interpreted without looking at the response of the opponent. Despite adaptation has been traditionally underestimated, or even disregarded, the fact that the two networks are competing for a common resource and with an objective that directly implies interaction and competition with other networks, suggests that new points of view coming from Network Science could be applied to understand and forecast the most adequate strategies (Aguirre et al., 2013).

Finally, it is possible to translate and generalize the results of network theory in football to organizational studies (Orlikowski, 1996; Padgett and Powell, 2012). When two teams (networks) are competing in the field, they need to develop strategies to create new options and "entrepreneurial actions" that generate "surprises" to the opponent; however, is this about surprises or spaces for serendipity? (Dew, 2009). Furthermore, the interaction between teams goes beyond the notion of "adaptation" challenging the concept of the "interface" or dynamic limit between the two teams. We believe that teams (beyond football and sports) need to generate new competencies such as systematic creativity and organizational learning that allow them to anticipate to the competition, promote their superiority and have more options to win and perform. With this regard, we should focus on how to create order and optimal organizational structures but, at the same time, to generate "disorder" in the opponent with the aim of generating situations of superiority.

## ACKNOWLEDGMENTS

J.M.B. is founded by MINECO (FIS2013-41057-P and FIS2017-84151-P). J.M.B. would like to thank C. Bielza, P. Cintia, M. Fernández, C. Fernández Conde, E. Granero, M. López, L. Pappalardo, C. Ramiro, F. Seirul.lo and J. Vilá for fruitful conversations about passing networks. J.H.M. would like to thank M. Chavez for his valuable comments. José L. Herrera Diestra is supported by the São Paulo Research Foundation (FAPESP) under grants 2016/01343-7 and 2017/00344-2.

## AUTHOR CONTRIBUTIONS

All authors participated in the conception of the article. J.M.B. wrote the initial draft. All authors revised the manuscript together. Datasets were provided by Opta.